\newcommand{\diracslash}[1]{#1\llap{/\kern2pt}}

\newcommand{\be}{\begin{equation}}
\newcommand{\ee}{\end{equation}}
\newcommand{\bea}{\begin{eqnarray}}
\newcommand{\eea}{\end{eqnarray}}
\newcommand{\ba}[1]{\begin{array}{#1}}
\newcommand{\ea}{\end{array}}

\newcommand{\bt}{\begin{tabular}}
\newcommand{\et}{\end{tabular}}

\newcommand{\beas}{\begin{eqnarray*}}
\newcommand{\eeas}{\end{eqnarray*}}

\documentclass[preprint,prd,aps,floats,nofootinbib,showpacs,floatfix]{revtex4}
\usepackage{graphicx}
\begin{document}

\title{Bottomonium states in hot asymmetric strange hadronic matter}
\author{Amruta Mishra}
\email{amruta@physics.iitd.ac.in,mishra@th.physik.uni-frankfurt.de}
\affiliation{Department of Physics, Indian Institute of Technology, Delhi,
Hauz Khas, New Delhi -- 110 016, India}

\author{Divakar Pathak}
\email{dpdlin@gmail.com}
\affiliation{Department of Physics, Indian Institute of Technology, Delhi,
Hauz Khas, New Delhi -- 110 016, India}

\begin{abstract}
We calculate the in-medium masses of the bottomonium states ($\Upsilon(1S)$,
$\Upsilon(2S)$, $\Upsilon(3S)$ and $\Upsilon(4S)$) in isospin asymmetric 
strange hadronic matter at finite temperatures. The medium modifications
of the masses arise due to the interaction of these heavy quarkonium states 
with the gluon condensates of QCD. The gluon condensates in the hot hadronic 
matter are computed from the medium modification of a scalar dilaton field
within a chiral SU(3) model, introduced in the hadronic model to incorporate 
the broken scale invariance of QCD. There is seen to be drop in the
masses of the bottomonium states and mass shifts are observed to be 
quite considerable at high densities for the excited states. 
The effects of density, isospin asymmetry, strangeness as well as 
temperature of the medium on the masses of the $\Upsilon$-states
are investigated. The effects of the isospin asymmetry
as well as strangeness fraction of the medium are seen to be appreciable
at high densities and small temperatures. The density effects are the most
dominant medium effects which should have observable consequences 
in the compressed baryonic matter (CBM) in the heavy ion collision 
experiments in the future facility at FAIR, GSI. The study of the
$\Upsilon$ states will however require access to energies higher 
than the energy regime planned at CBM experiment. The density effects
on the bottomonium masses should also show up in the dilepton spectra
at the SPS energies, especially for the excited states for which the 
mass drop is observed to quite appreciable.

\end{abstract}
\maketitle

\def\bfm#1{\mbox{\boldmath $#1$}}

\section{Introduction}
The study of properties of hadrons in hot and dense matter is an important
topic of contemporary research in strong interaction physics. The medium 
modifications of the hadrons affect the experimental observables from the 
strongly interacting matter created in the relativistic heavy ion collision 
experiments. The in-medium properties of the light vector mesons
\cite{wambach,hatsuda,klinglnpa} are of 
relevance for observables like dilepton spectra, whereas modifications 
of the kaons and antikaons modify the production and propagation of 
these particles. The modifications of the masses of the charm mesons, 
$D$ ($\bar D$) mesons as well as of the charmonium states in the medium 
can modify the yield of the open charm mesons and of the charmonium states. 
A larger drop of the masses of the $D$ mesons as compared to the mass drop
of the charmonium states can open decay channels of the excited states
of charmonium to $D\bar D$ in the medium, which are not accessible 
in vacuum. This can reduce the production of $J/\psi$ from the decay of
these excited charmonium states and thereby lead to J/$\psi$ suppression 
in heavy ion collision experiments.  In high energy nuclear collisions, 
$J/\psi$ suppression could also be due to formation of
quark gluon plasma \cite{blaiz,satz}. 
In the presence of quark gluon plasma, the quark antiquark potential 
is screened leading to decrease in the binding energy 
of the quarkonium states and when the distance between
the quark and antiquark becomes larger than the  inverse of the
Debye mass, the $Q\bar Q$ can no longer exist as a bound state.
This can lead to step-wise melting of the quarkonium states
from the hot deconfined matter and hence cause suppression of
the ground states $J/\psi$ and $\Upsilon(1S)$ as the feed-down 
from the excited quarkonium states will no longer be possible
at temperatures above the dissociation temperatures of the
excited states \cite{satz2013}. At still higher temperatures,
the quarkonium ground state can also dissociate due
to color screening. The suppression of the heavy 
quarkonium states in the deconfined matter \cite{mocsy2013}, 
can also be due to the
processes of gluo-dissociation \cite{satzgluodiss} and 
inelastic parton scattering \cite{rapppartscatt}.
In the former process, the quarkonium state breaks up due to
interaction with a hard thermal gluon which changes the color singlet
quarkonium state to an unbound color octet $Q\bar Q$ pair.
In the inelastic parton scattering, the quarkonium state 
interacts with a parton (quark or gluon) through exchange 
of gluons, leading to dissociation of the heavy quarkonium.   
An effective field theory describing the quarkonia systems
through potentials, namely pNRQCD (potential nonrelativistic QCD)
has been extensively used in the literature \cite{brambilla1}. 
Within this framework, two mechanisms
contributing to the decay width of the quarkonium were
identified as singlet-to-octet thermal breakup
and Landau damping. In the leading order, these correspond
to gluo-dissociation and dissociation of quarkonium due
to inelastic parton scattering \cite{brambilla2}.

The masses of the $D$ ($\bar D$) mesons, which consist of a heavy charm 
quark (antiquark) and a light antiquark (quark), within the QCD sum rule 
approach, are modified largely due to their interaction with the light quark 
condensates in the hadronic medium \cite{haya1}. On the contrary, the masses of 
the heavy quarkonium states, e.g., the charmonium states, are modified
in the leading order due to their interaction with the gluon 
condensates in the hadronic medium \cite{leeko,leeko2,kimlee}. 
These gluon condensates can be written in terms of the color electric and 
color magnetic fields. The modifications of the masses of the charmonium 
states, due to changes in the gluon condensates in the nuclear medium 
have been studied in the linear density approximation \cite{leeko,leeko2}, 
using the leading order QCD formula \cite {pes1}. 
The formula for the charmonium mass shift becomes proportional to 
$\langle \frac{\alpha_s}{\pi} {\vec E}^2\rangle$, 
similar to the second order Stark effect \cite{leeko,leeko2}, since
the Wilson coefficient for the operator $\langle \frac{\alpha_s}{\pi} 
{\vec B}^2\rangle$ vanishes in the non-relativistic limit. 
These studies \cite{leeko,leeko2} show the drop of the $J/\psi$ mass 
at the nuclear matter saturation density to be quite small, whereas 
the masses of the excited charmonium states, e.g. $\psi(3686)$ and 
$\psi(3770)$ are observed to have appreciable drop in the nuclear medium. 
The charmonia masses in hot strange hadronic matter  have been studied 
using the leading order QCD formula,
due to changes in the gluon condensates in the hadronic medium
calculated in a chiral effective model \cite{amepja}. The gluon
condensate of QCD is simulated by a scalar dilaton field introduced 
in the effective hadronic model to incorporate scale symmetry breaking 
of QCD. The medium modifications of the decay widths of the charmonium 
states to $D\bar D$ have been studied arising from the mass shifts 
of the $D$ mesons in the medium  \cite{frimanlee} as well as
of the charmonium states \cite{amepja} accounting for the internal 
quark structure of the charmonium as well as $D(\bar D)$ mesons
within the so called $^3P_0$ model. The charmonium decay widths in the medium 
have also been studied recently using a field theoretical model 
for composite hadrons with constituent quarks \cite{amspmchm}.
Within the QCD sum rule approach, the mass modifications of 
the charmonium states, $J/\psi$ as well as $\eta_c$ have  also been 
studied \cite{amchqsr} as arising from the medium modifications of the
scalar gluon condensate and the twist-2 gluon operator, computed from
the medium modification of the dilaton field in the chiral effective model.
The mass shift of the D meson at finite density was studied using the QCD sum
rule approach \cite{haya1} using condensates upto dimension 4 in the operator
product expansion and the dominant contribution to the D meson mass drop comes
from the term $m_c\langle \bar q q \rangle$, which turns out to be appreciable
due to the large value of $m_c$. 
The mass shifts and the splitting of the $D$-$\bar D$ (as well as
$B$-$\bar B$) meson masses have been studied within the QCD sum rule approach 
by considering condensate operators upto dimension 5 in the operator product
expansion. This includes also the contributions from the gluon condensates 
as well as the mixed quark-gluon condensate \cite{hilgerprc,hilgerjpg},
with the dominant contribution to the mass modifications of the $D(B)$ mesons 
arising from the term $m_c \langle \bar q q \rangle$ 
($m_b \langle \bar q q \rangle$) in the operator product expansion.
In the recent past, the  discovery of many hadrons including the heavy charm
\cite{physreplee2010} and bottom quarks, as well as, the observed suppression 
of the heavy quarkonia at SPS, RHIC and LHC, which could indicate
the formation of deconfined matter, has added further motivation to study 
the properties of the heavy quarkonium states in the hot and dense matter
resulting from heavy ion collision experiments. 
The charmonium suppression has been observed at SPS energies
\cite{spsexpt} and at RHIC \cite{rhicjpsisupp}.
The bottomonium suppression has been observed at RHIC
for Au+Au collision at $\sqrt s$=200 GeV \cite{starcoll}
as well as at LHC in Pb+Pb collision at $\sqrt s$= 2.76 TeV
\cite{cms}. The quarkonia production have
been studied using a rate equation \cite{rappchmbott}
for SPS (for Pb-Pb collision at 158 AGeV ($\sqrt {s}$=17.3 GeV)), 
RHIC (Au+Au collisions at $\sqrt s$=200 GeV) and LHC 
(Pb+Pb collision at $\sqrt s$=2.76 TeV) energies, 
including their production from initial nucleon nucleon 
hard scattering and their regeneration from produced QGP, 
as well as accounting for the in-medium effects of quarkonia
\cite{kochmbott}.
At RHIC and LHC, the medium effects on the quarkonia production
are due to the thermal effects, whereas the density effects
on the heavy quarkonia production should be important at SPS, CERN
as well as at the Compressed baryonic matter (CBM) experiment
at FAIR project at the future facility at GSI
\cite{gsifuture}.
At the CBM experiment at FAIR, one of the areas
of focus for research will be the study of rare probes, e.g.,
the heavy quarkonia. The measurements of the production of the
heavy quarkonia will be possible due to the high beam intensities
and long running times to be used in these experiments.
The experimental facility will be using heavy ion beams
in fixed target mode with beam energy of about 10 to 45 AGeV
($\sqrt s$=4.5 to 10 GeV). These energies will make the
study of charmonia possible quite extensively, whereas, a study 
of bottomonia production will require access to higher energies,
as the top energy of CBM is about the threshold energy
when $b$ and $\bar b$ can be pair produced which can later 
combine to form the bottomonium state.
The density effects, which seem to the important medium 
effects on the bottomonia masses in the present investigation,
should also be possible to study at SPS energies,
where the yields of the $\Upsilon$ states have already been 
measured in pA collisions at incident energy of 450 AGeV
($\sqrt s$=29.1GeV) by the NA50 Collaboration \cite{pANA50bottomonium}.
These measurements can provide a baseline for the $\Upsilon$ 
production studies to be carried out in the ion-ion collisions
at higher centre of mass energies. 

In the present work, we study the in-medium masses of the bottomonium states,
$\Upsilon(1S)$, $\Upsilon(2S)$, $\Upsilon(3S)$ and $\Upsilon(4S)$
in hot asymmetric strange hadronic matter due to the interaction 
with the gluon condensates using the leading order QCD formula. 
The medium modification of the gluon condensate in the hadronic
medium is calculated from the medium modification of a scalar 
dilaton field introduced within a chiral SU(3) model \cite{papa}
to incorporate scale symmetry breaking of QCD.
The model has been used to study the in-medium properties
of the vector mesons \cite{hartree,kristof1}, kaons and antikaons 
\cite{isoamss,isoamss2}. The model has then been generalized to chiral SU(4) 
to derive the interactions of the charm  $D(\bar D)$ mesons with the light 
hadron sector and study the effects of isospin asymmetry \cite{amarind},
strangeness \cite{amepja} and temperature \cite{amdmeson}
on the mass modifications of these mesons in the hadronic medium. 
The mass shifts of the charmonium states have been calculated due to the
changes in the gluon condensates in the medium \cite{amepja,amchqsr} 
obtained from the change of a scalar dilaton field, which mimics 
the scale symmetry breaking of QCD, in the chiral effective model. 
In the present investigation, we study the in-medium masses
of bottomonium states, obtained from the dilaton field, $\chi$,
calculated for the asymmetric strange hadronic matter at finite 
temperatures.

The outline of the paper is as follows : In section II, we discuss 
briefly the chiral $SU(3)$ model which has been used to investigate 
the mass modification of the bottomonium states in the present work. 
The medium modifications of the bottomonium masses arise from the medium 
modification of a scalar dilaton field introduced in the hadronic 
model to incorporate broken scale invariance of QCD leading to QCD 
trace anomaly. In section III, we summarize the results obtained 
in the present investigation.

\section{The hadronic chiral $SU(3) \times SU(3)$ model}
We use a chiral $SU(3)$ model \cite{papa,weinberg,coleman,bardeen}, 
which incorporates the scale symmetry breaking of QCD through introduction 
of a scalar dilaton field \cite{sche1,ellis} within the hadronic model. The
modification of the gluon condensates in the hadronic matter is obtained
from the medium modification of the scalar dilaton field, which then
is used to investigate the in-medium masses of the bottomonium states
in the hot isospin asymmetric strange hadronic matter. 
The effective hadronic chiral Lagrangian density contains 
the following terms:
\begin{equation}
{\cal L} = {\cal L}_{kin}+\sum_{W=X,Y,V,A,u} {\cal L}_{BW} + 
{\cal L}_{vec} + {\cal L}_{0} + {\cal L}_{scalebreak}+ {\cal L}_{SB}
\label{genlag}
\end{equation}
In Eq. (\ref{genlag}), ${\cal L}_{kin}$ is kinetic energy term, 
${\cal L}_{BW}$ is the baryon-meson interaction term in which the 
baryon-spin-0 meson interaction term generates the vacuum baryon masses. 
${\cal L}_{vec}$  describes the dynamical mass generation of the vector 
mesons via couplings to the scalar mesons and contain additionally 
quartic self-interactions of the vector fields. ${\cal L}_{0}$ contains 
the meson-meson interaction terms inducing the spontaneous breaking of 
chiral symmerty, ${\cal L}_{scalebreak}$ is the scale invariance breaking 
logarithmic term. ${\cal L}_{SB}$ describes the explicit chiral symmetry breaking. 
To study the hadron properties at finite temperature and densities
in the present investigation, we use the mean  field approximation,
where all the meson fields are treated as classical fields. 
In this approximation, only the scalar and the vector fields 
contribute to the baryon-meson interaction, ${\cal L}_{BW}$
since for all the other mesons, the expectation values are zero.
The scale breaking term \cite{sche1,ellis} of the Lagrangian density
\begin{equation}
{\cal L}_{scalebreak}= -\frac{1}{4} \chi^{4} 
{\rm ln} \frac{\chi^{4}}{\chi_{0}^{4}} + \frac{d}{3} \chi^{4} 
{\rm ln} \Bigg( \frac{\left( \sigma^{2} - \delta^{2}\right)\zeta }
{\sigma_{0}^{2} \zeta_{0}} \Big( \frac{\chi}{\chi_{0}}\Big) ^{3}\Bigg), 
\label{scalebreak}
\end{equation}
where $\chi$, $\sigma$, $\zeta$ and $\delta$ are the scalar dilaton field,
non-strange scalar field, strange scalar field and the scalar-isovector
field respectively.
The effect of these logarithmic terms is to break the scale invariance, 
which leads to the trace of the energy momentum tensor as \cite{heide1}
\begin{equation}
\theta_{\mu}^{\mu} = \chi \frac{\partial {\cal L}}{\partial \chi} 
- 4{\cal L} 
= -(1-d)\chi^{4}.
\label{tensor1}
\end{equation}

The trace of the energy momentum tensor
in QCD  is given as
\begin{equation}
\theta_{\mu}^{\mu} = \langle \frac{\beta_{QCD}}{2g} 
G_{\mu\nu}^{a} G^{\mu\nu a} \rangle + \sum_i m_i \bar {q_i} q_i ,
\label{tensorquark}
\end{equation}
where the second term in the trace accounts for the finite 
quark masses, with $m_i$ as the current quark mass for the quark
of flavor, $i=u,d,s$. 
The one loop QCD $\beta$ function is given as
\begin{equation}
\beta_{\rm {QCD}} \left( g \right) = -\frac{11 N_{c} g^{3}}{48 \pi^{2}} 
\left( 1 - \frac{2 N_{f}}{11 N_{c}} \right),
\label{qcdbeta}
\end{equation}
where $N_c=3$ is the number of colors and $N_f$ is the number of
quark flavors.
Comparing the trace of the energy momentum tensor in the chiral
effective model given by (\ref{tensor1}) to that of QCD given by 
(\ref{tensorquark}) and using equation (\ref{qcdbeta}), 
we obtain the scalar gluon condensate related to the dilaton field 
as,
\begin{equation}
\langle \frac{\alpha_{s}}{\pi} 
G_{\mu\nu}^{a} G^{\mu\nu a} \rangle = \frac{24}{(33-2N_f)}\left[ (1-d)\chi^{4} 
+ \sum_i m_i \bar {q_i} q_i\right].
\label{chiglu}
\end{equation} 
The second term, $\sum_i m_i \bar q_i q_i$
can be identified to be the negative of the explicit chiral 
symmetry breaking term ${\cal L}_{SB}$ of equation (\ref{genlag})
\cite{amchqsr} and is given as
\begin{eqnarray}
\sum_i m_i \bar {q_i} q_i &=& - {\cal L} _{SB}  =  
\Big[ m_{\pi}^{2} 
f_{\pi} \sigma 
+ \left( \sqrt{2} m_{k}^{2}f_{k} - \frac{1}{\sqrt{2}} 
m_{\pi}^{2} f_{\pi} \right) \zeta 
\Big].
\label{lsb}
\end{eqnarray}
We thus see from the equation (\ref{chiglu}) that the scalar 
gluon condensate $\left\langle \frac{\alpha_{s}}{\pi} G_{\mu\nu}^{a} 
G^{\mu\nu a}\right\rangle$ is proportional to the fourth power of the 
dilaton field, $\chi$, in the limiting situation of massless quarks
\cite{amepja}.

The coupled equations of motion for the non-strange scalar field $\sigma$, 
strange scalar field $ \zeta$, scalar-isovector field $ \delta$ and dilaton 
field $\chi$, are derived from the Lagrangian density.
These equations of motion are solved to obtain the density 
and temperature dependent values of the scalar fields ($\sigma$,
$\zeta$ and $\delta$) and the dilaton field, $\chi$, in the isospin
asymmetric hot strange hadronic medium. The values of the scalar fields 
and the dilaton field, for baryon density, $\rho_B$ and temperature, T,
are calculated for given values of the strangeness fraction of the medium, 
$f_s= \frac{\sum_{i} s_{i} \rho_{i}}{\rho_{B}}$, and isospin asymmetry parameter, 
$\eta= -\frac{\sum_{i}  I_{3i} \rho_{i}}{\rho_{B}}$, where $s_{i}$ 
is the number of strange quarks of baryon $i$ (i=p,n,$\Lambda$,
$\Sigma ^{\pm,0}$, $\Xi^{-,0}$) and $I_{3i}$ is the third component of  
isospin of the $i$-th baryon.
The in-medium masses of heavy quarkonium states are 
modified due to the modifications of the  scalar gluon and the twist-2
gluon condensates \cite{haya1,leeko} in the hadronic medium. These gluon
condensates can be written in terms of the color electric and
color magnetic fields, $\langle \frac{\alpha_s}{\pi} {\vec E}^2\rangle$ 
and $\langle \frac{\alpha_s}{\pi} {\vec B}^2\rangle$ \cite{david} and 
as has already been mentioned, for heavy quarkonium states, the contribution 
from the magnetic field part vanishes.
Hence the mass shift for these states in the hadronic medium arise due
to the change in the electric field part, $\frac{\alpha_s}{\pi} 
\langle {\vec E}^2\rangle$, similar to the second order Stark effect 
\cite{leeko}. In the leading order mass shift formula derived in the large 
bottom quark mass limit \cite{pes1}, the shift in the mass of the bottomonium
state is given as
\cite{leeko}
\begin{equation}
\Delta m_{\Upsilon} = -\frac{1}{9} \int dk^{2} \vert 
\frac{\partial \psi (k)}{\partial k} \vert^{2} \frac{k}{k^{2} 
/ m_{b} + \epsilon} \bigg ( 
\left\langle  \frac{\alpha_{s}}{\pi} E^{2} \right\rangle-
\left\langle  \frac{\alpha_{s}}{\pi} E^{2} \right\rangle_{0}
\bigg ).
\label{massl}
\end{equation}
In the above, $m_b$ is the mass of the bottom quark,
$m_\Upsilon$ is the vacuum mass of the bottomonium state 
and $\epsilon = 2 m_{b} - m_{\Upsilon}$ represents the binding energy
of the bottomonium state. 
$\psi (k)$ is the wave function of the bottomonium state
in the momentum space, normalized as $\int\frac{d^{3}k}{(2\pi)^{3}} 
\vert \psi(k) \vert^{2} = 1 $ \cite{leetemp}.
It may be noted here that the leading order perturbative formula
(\ref{massl}) for the quarkonium ($Q\bar Q$) state 
has been derived \cite{pes1} in the limit of large quark 
mass, so that the inverse of the radius of the quarkonium state 
is much larger than its binding energy, $\epsilon$. 
Furthermore, the  $Q\bar Q$ binding
in the heavy quarkonium state is approximated to be
one-gluon exchange Coulomb potential, where the gluon energy is
small compared to the binding energy of the quarkonium state. 
The wave function of the quarkonium state in this approximation
is Coulombic. However, the Coulombic wave function for the
quarkonium state may not be realistic for the excited states 
of the quarkonium state \cite{leeko}.
The mass modifications of the charmonium states due to their 
interaction with the gluon condensates were studied
in Ref. \cite{leeko,amepja}, assuming the wave functions to be 
harmonic oscillator type. In the present work, we also use the
harmonic oscillator wave functions to study of the medium 
modifications of the masses of the bottomonium states,
due to their interactions with the gluon condensates.  

In the non-relativistic limit, due to vanishing of the 
contribution from the magnetic field,
the expectation value of the scalar gluon condensate can be expressed 
in terms of the color electric field and hence the mass shift formula
for the bottomonium states can be written in terms of the difference
in the value of the scalar gluon condensate in the medium and in the vacuum
as \cite{amepja} 
\begin{eqnarray}
\Delta m_{\Upsilon} &=& \frac{1}{18} \int dk^{2} \vert 
\frac{\partial \psi (k)}{\partial k} \vert^{2} \frac{k}{k^{2} 
/ m_{b} + \epsilon} 
\bigg ( 
\left\langle \frac{\alpha_{s}}{\pi} 
G_{\mu\nu}^{a} G^{\mu\nu a}\right\rangle -
\left\langle \frac{\alpha_{s}}{\pi} 
G_{\mu\nu}^{a} G^{\mu\nu a}\right\rangle _{0}
\bigg ).
\label{mass1}
\end{eqnarray}

In the present investigation, the wave functions for the bottomonium states 
are taken to be Gaussian and are given as \cite{frimanlee}
\begin{equation}
\psi_{N, l} = {\tilde {\rm N}} Y_{l}^{m} (\theta, \phi) 
(\beta^{2} r^{2})^{\frac{l}{2}} e^{-\frac{1}{2} \beta^{2} r^{2}} 
L_{N - 1}^{l + \frac{1}{2}} \left( \beta^{2} r^{2}\right)
\label{wavefn} 
\end{equation} 
where $\beta^{2} = M \omega / \hbar$ is the parameter describing the strength 
of the harmonic oscillator potential, $M = m_{b}/2$ is the reduced mass of 
the bottom quark and bottom anti-quark system, $L_{p}^{k} (z)$ 
is the associated Laguerre polynomial. $\tilde {\rm N}$ is the normalization constant
determined from $\int |\psi_{N,l}({\bf r})|^2 d^3 r=1$. 
The oscillator constant $\beta$ is calculated from the decay widths
of the bottomonium state to $e^+e^-$ given by the formula 
\cite{vanroyen,spmmesonspect}

\begin{equation}
\Gamma_{\Upsilon \rightarrow e^+e^-}=\frac{16 \pi \alpha^2}{9M_\Upsilon^2}
|\psi (\bf 0)|^2,
\label{leptonicdw}
\end{equation}
where, $\alpha=\frac{1}{137}$ is the fine structure constant and $\psi(\bf 0)$ 
is the wave function of the bottomonium state at the origin.
Knowing the wave functions of the bottomonium states and 
calculating the medium modification of the scalar gluon condensate from
the in-medium dilaton field and the scalar fields,
we obtain the mass shift of the bottomonium states.
In the next section we discuss the results for these in-medium bottomonium 
masses in hot asymmetric strange hadronic matter obtained in the present work.

\section{Results and Discussions}

\begin{figure}
\includegraphics[width=16cm,height=16cm]{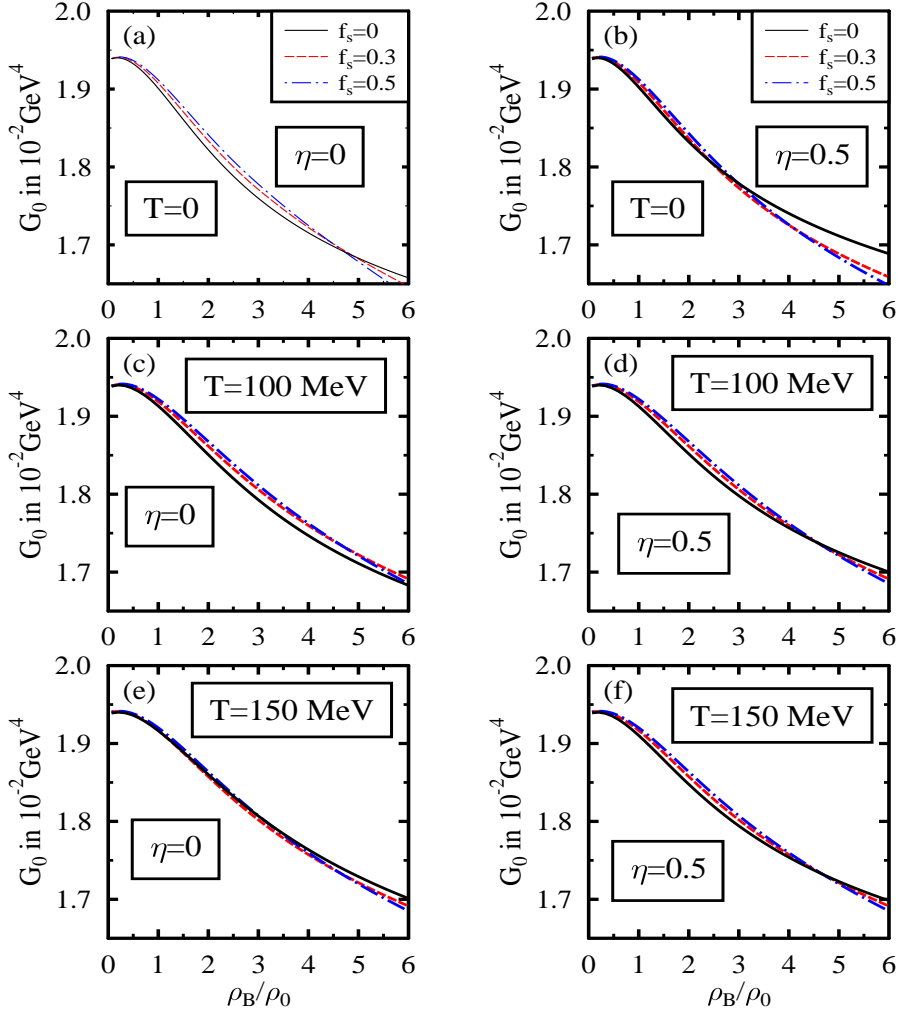}
\caption{(Color online)
The scalar gluon condensate, 
$G_0\equiv\langle (\alpha_s/{\pi})G_{\mu \nu}^aG^{\mu \nu a}\rangle$ 
is plotted as a function of the baryon density, $\rho_B$ in units of nuclear matter
saturation density, $\rho_0$. This is shown for different values of the
temperature for typical values of the strangeness fraction of the medium,
$f_s$, in subplots (a), (c) and (e) for the isospin symmetric matter   
and in subplots (b), (d) and (f) for the isospin asymmetric matter
with $\eta$=0.5.
}
\label{ggscdens}
\end{figure}

\begin{figure}
\includegraphics[width=16cm,height=16cm]{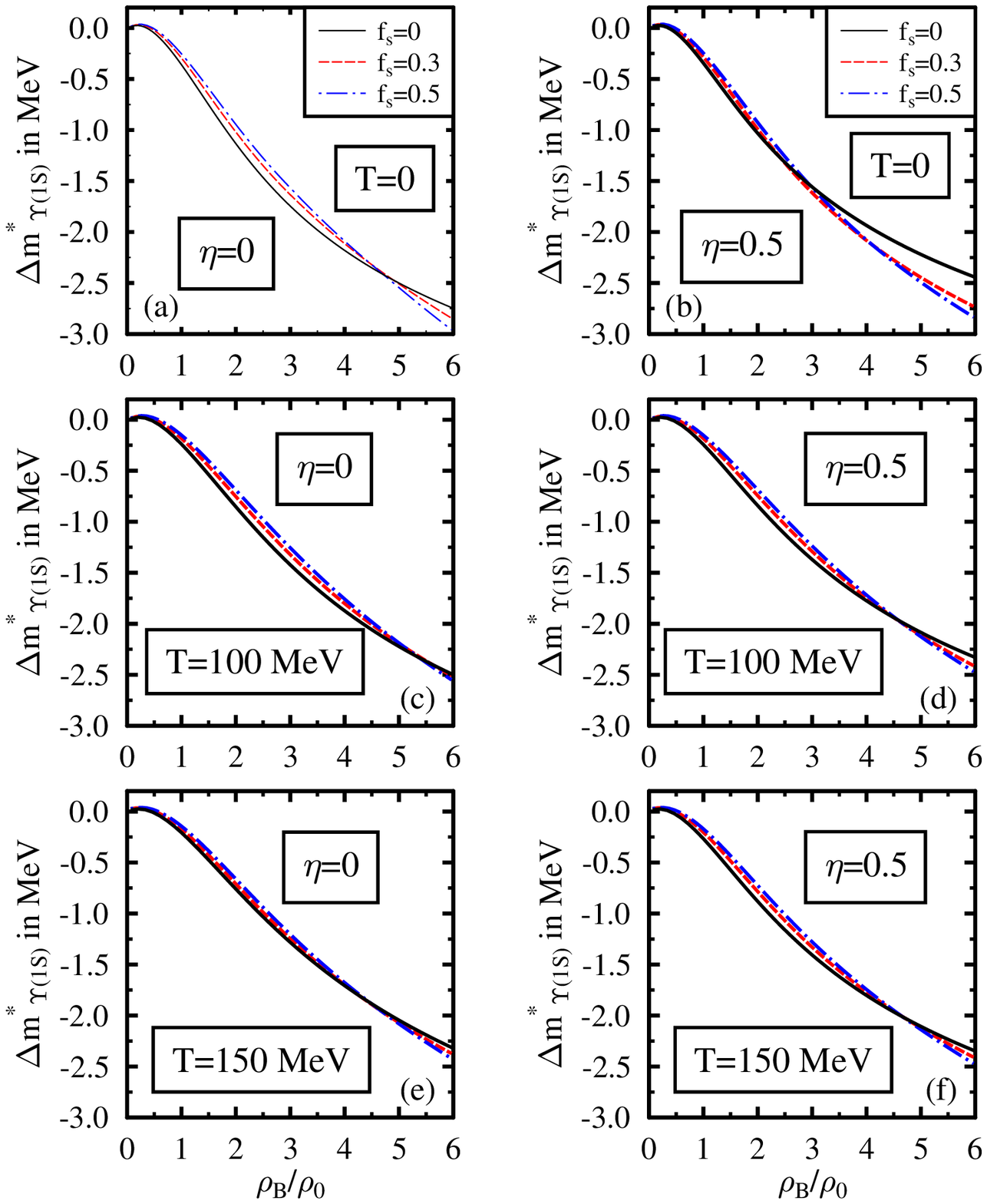}
\caption{(Color online)
The mass shift of $\Upsilon(1S)$ plotted as a function of the
baryon density in units of nuclear matter saturation density
at given temperatures,
for different values of the 
strangeness fraction for the isospin symmetric ($\eta$=0) as well as isospin 
asymmetric ($\eta$=0.5) hadronic matter.
}
\label{dmupsln1s}
\end{figure}

\begin{figure}
\includegraphics[width=16cm,height=16cm]{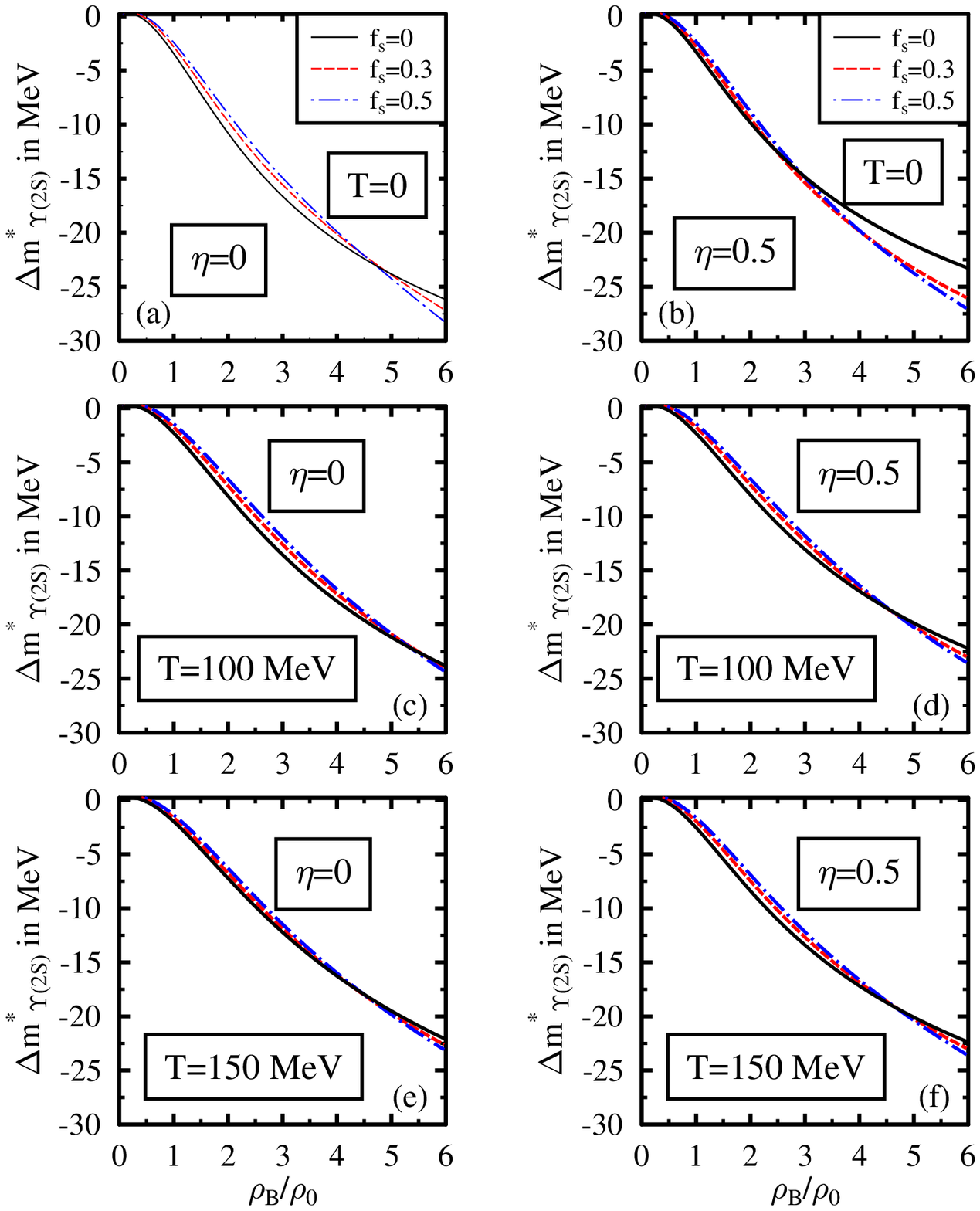}
\caption{(Color online)
The mass shift of $\Upsilon(2S)$ plotted as a function of the
baryon density in units of nuclear matter saturation density
at given temperatures,
for different values of the 
strangeness fraction for the isospin symmetric ($\eta$=0) as well as isospin 
asymmetric ($\eta$=0.5) hadronic matter.
}
\label{dmupsln2s}
\end{figure}

\begin{figure}
\includegraphics[width=16cm,height=16cm]{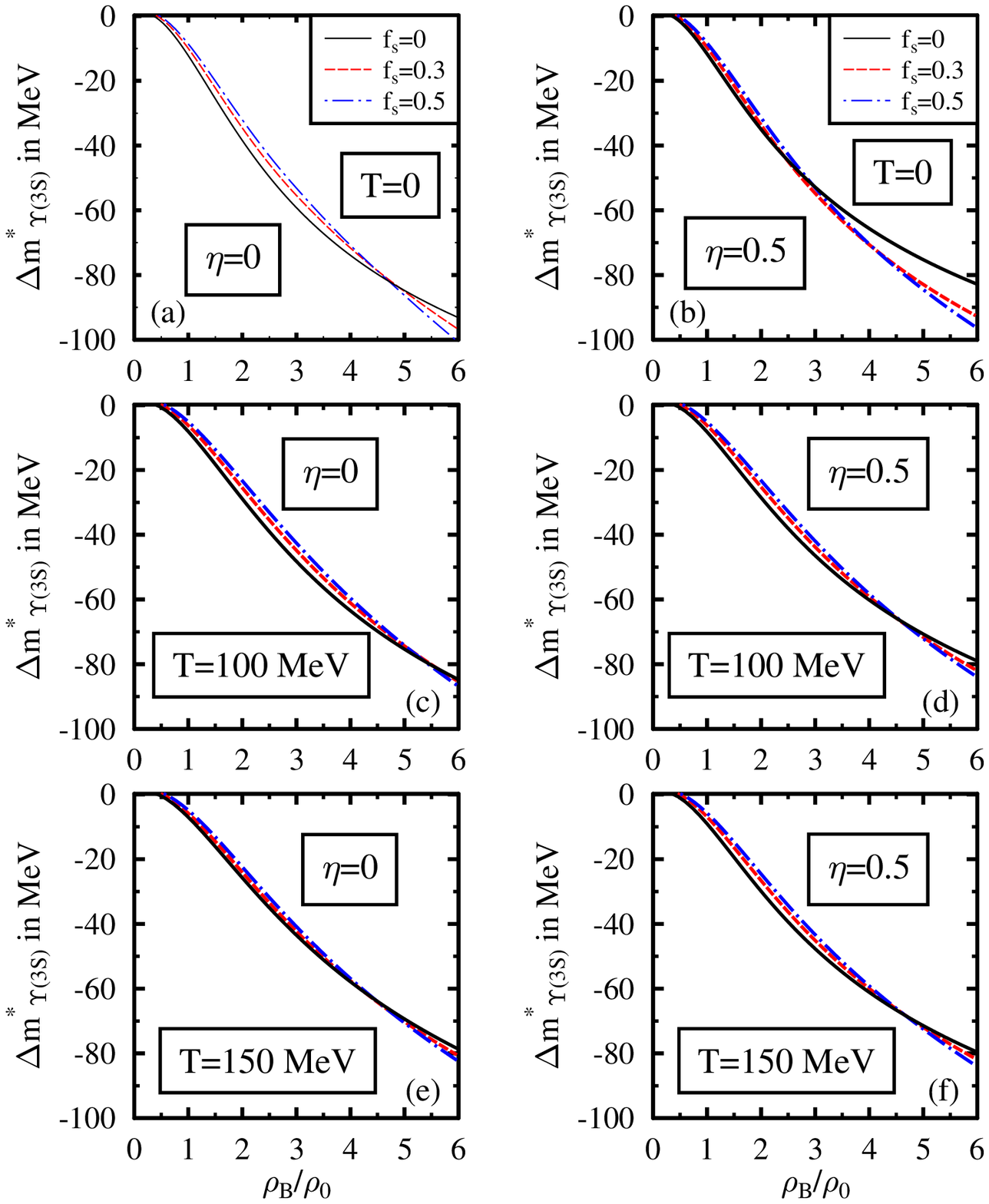}
\caption{(Color online)
The mass shift of $\Upsilon(3S)$ plotted as a function of the
baryon density in units of nuclear matter saturation density
at given temperatures,
for different values of the 
strangeness fraction for the isospin symmetric ($\eta$=0) as well as isospin 
asymmetric ($\eta$=0.5) hadronic matter.
}
\label{dmupsln3s}
\end{figure}

\begin{figure}
\includegraphics[width=16cm,height=16cm]{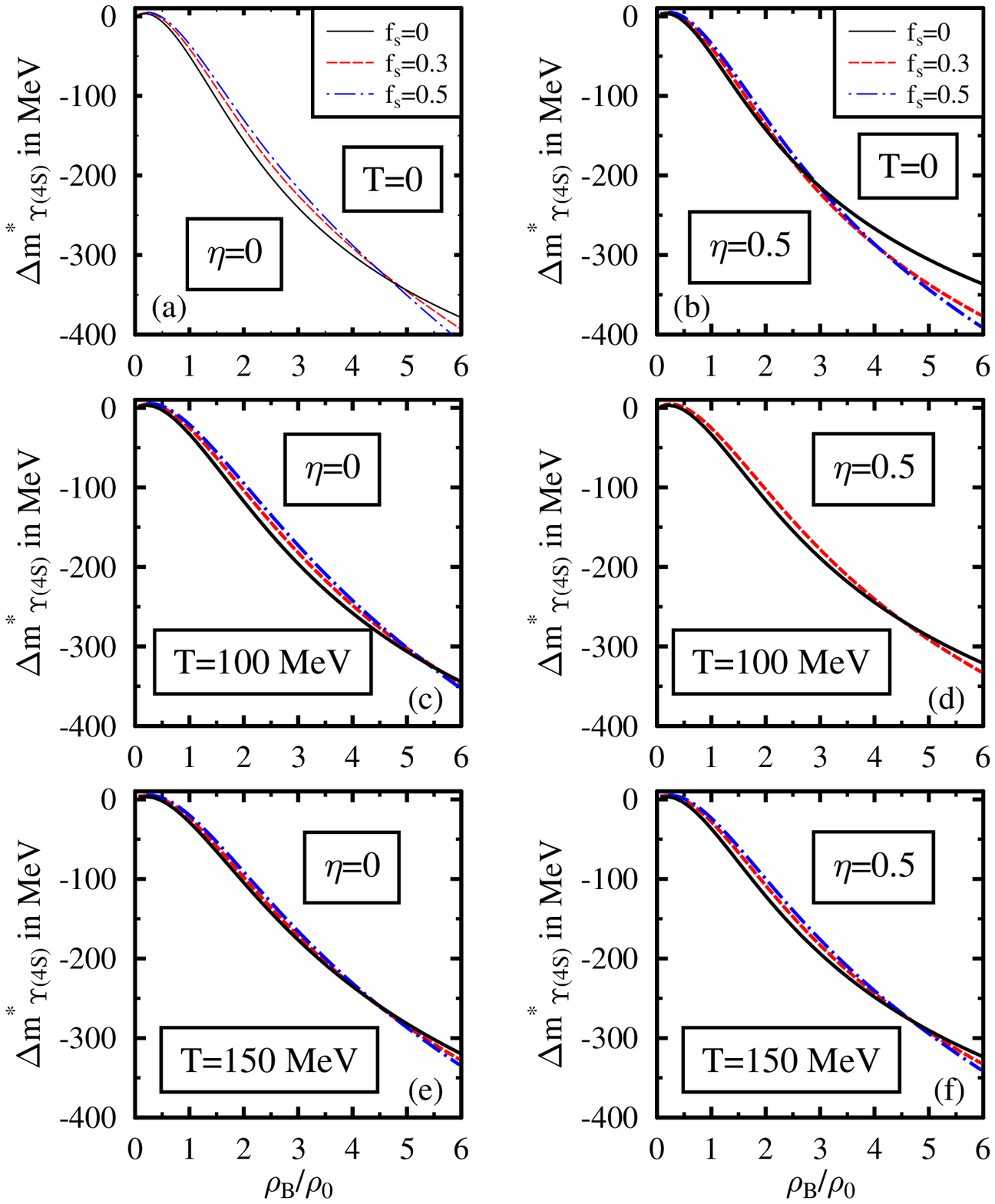}
\caption{(Color online)
The mass shift of $\Upsilon(4S)$ plotted as a function of the
baryon density in units of nuclear matter saturation density
at given temperatures, for different values of the 
strangeness fraction for the isospin symmetric ($\eta$=0) 
as well as isospin asymmetric ($\eta$=0.5) hadronic matter.
}
\label{dmupsln4s}
\end{figure}

\begin{figure}
\includegraphics[width=16cm,height=16cm]{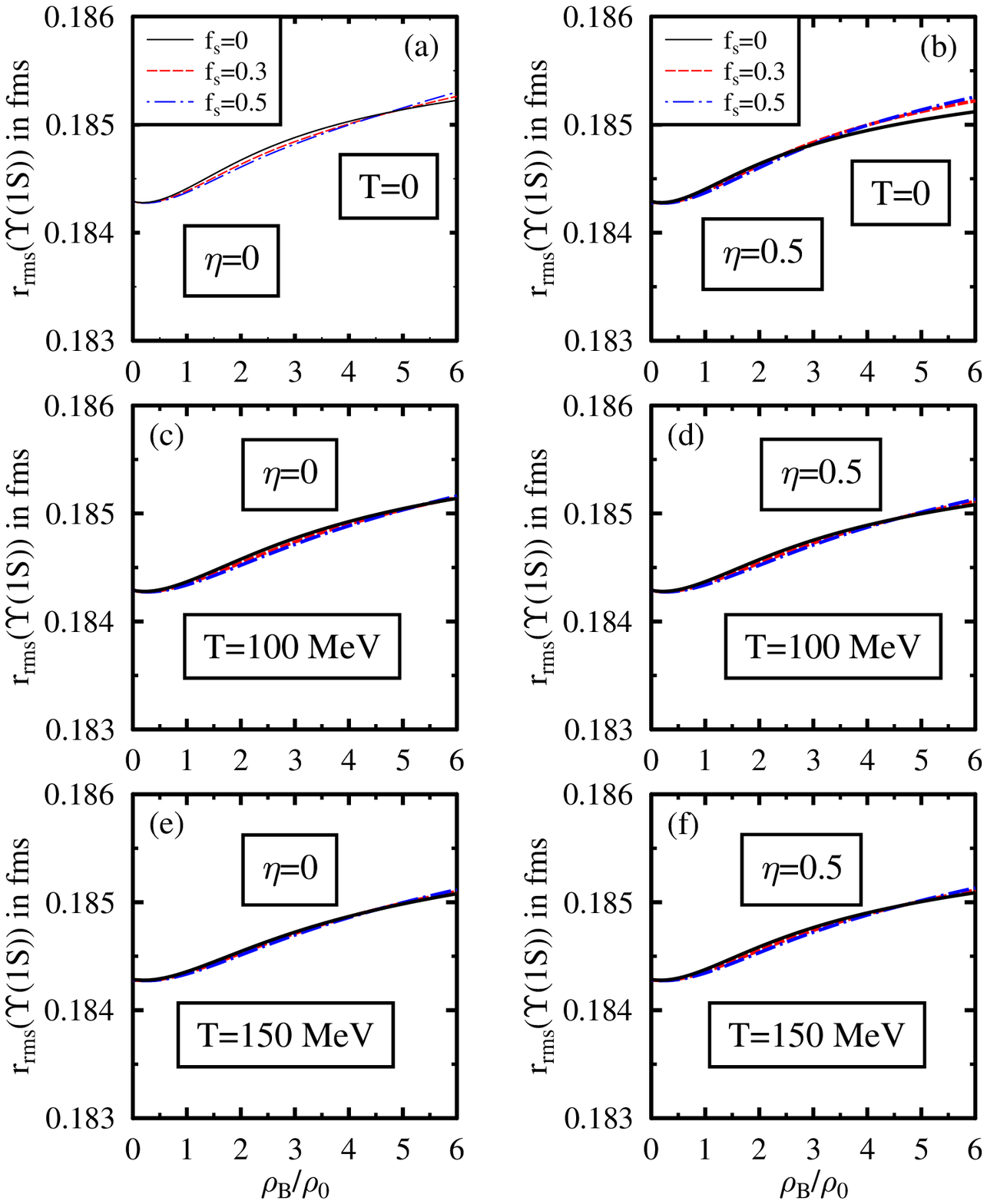}
\caption{(Color online)
The rms radius of $\Upsilon(1S)$ plotted as a function of the
baryon density in units of nuclear matter saturation density
at given temperatures,
for different values of the 
strangeness fraction for the isospin symmetric ($\eta$=0) as well as isospin 
asymmetric ($\eta$=0.5) hadronic matter.
}
\label{rmsupsln1s}
\end{figure}

\begin{figure}
\includegraphics[width=16cm,height=16cm]{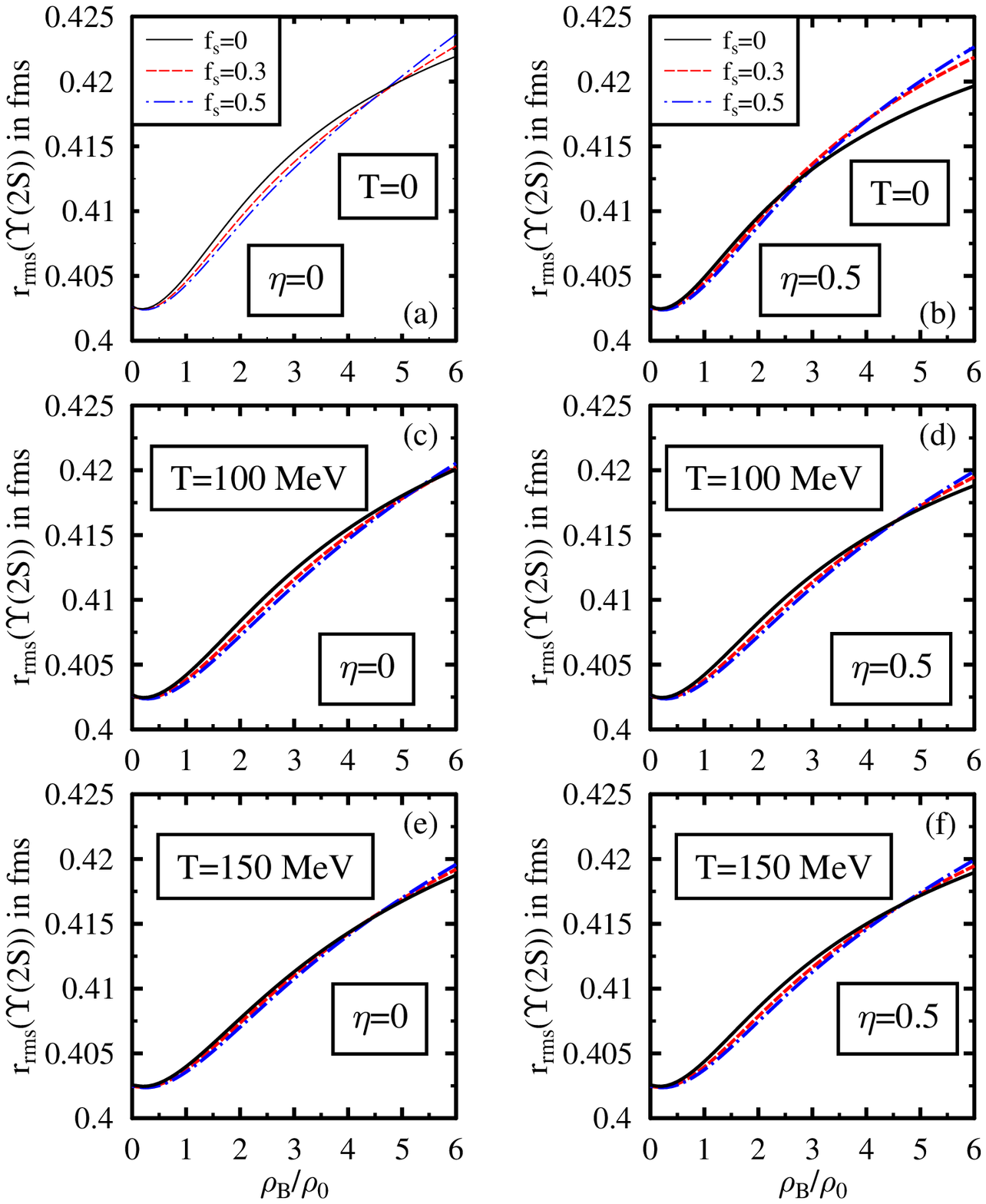}
\caption{(Color online)
The rms radius of $\Upsilon(2S)$ plotted as a function of the
baryon density in units of nuclear matter saturation density
at given temperatures,
for different values of the 
strangeness fraction for the isospin symmetric ($\eta$=0) as well as isospin 
asymmetric ($\eta$=0.5) hadronic matter.
}
\label{rmsupsln2s}
\end{figure}

\begin{figure}
\includegraphics[width=16cm,height=16cm]{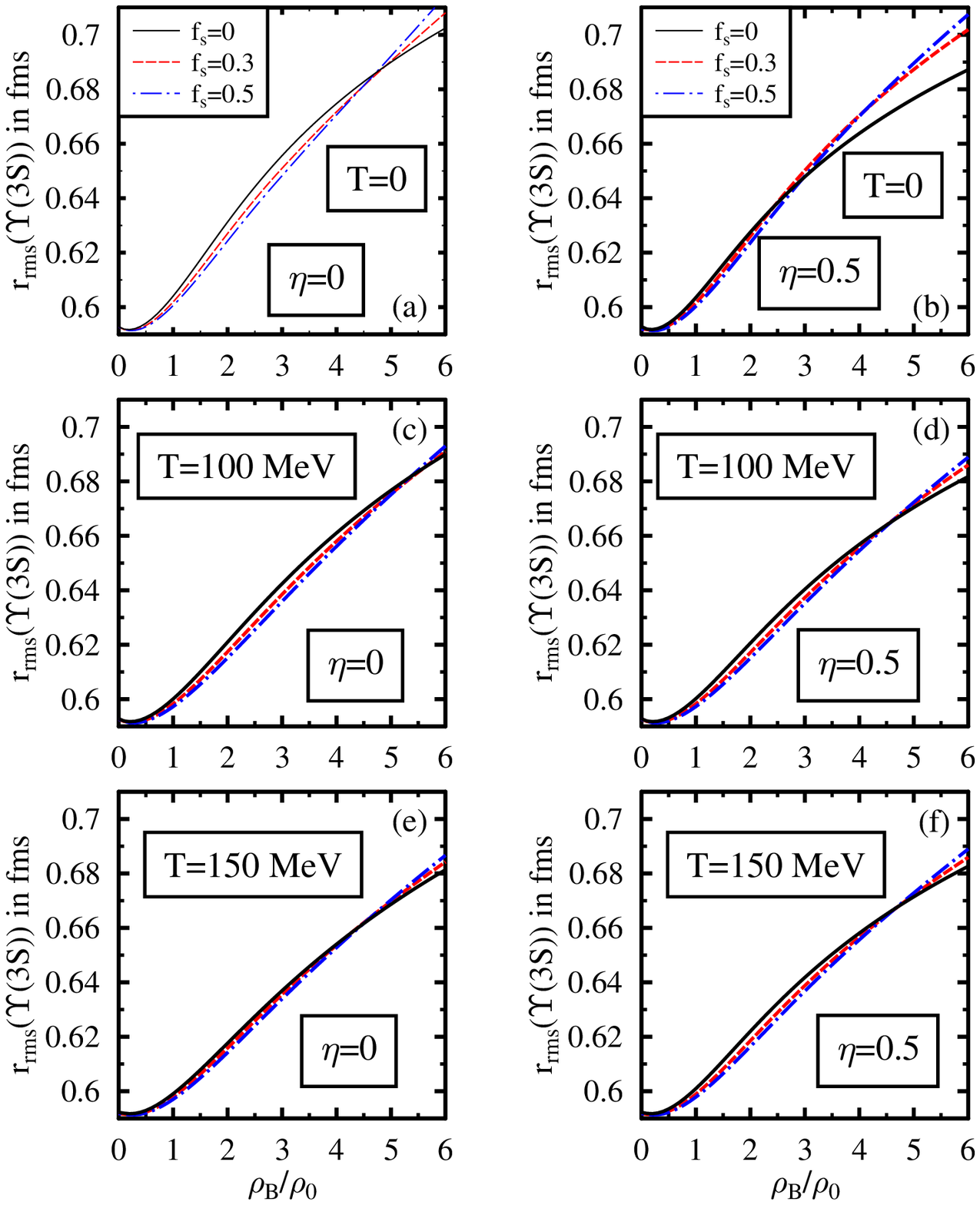}
\caption{(Color online)
The rms radius of $\Upsilon(3S)$ plotted as a function of the
baryon density in units of nuclear matter saturation density
at given temperatures,
for different values of the 
strangeness fraction for the isospin symmetric ($\eta$=0) as well as isospin 
asymmetric ($\eta$=0.5) hadronic matter.
}
\label{rmsupsln3s}
\end{figure}

\begin{figure}
\includegraphics[width=16cm,height=16cm]{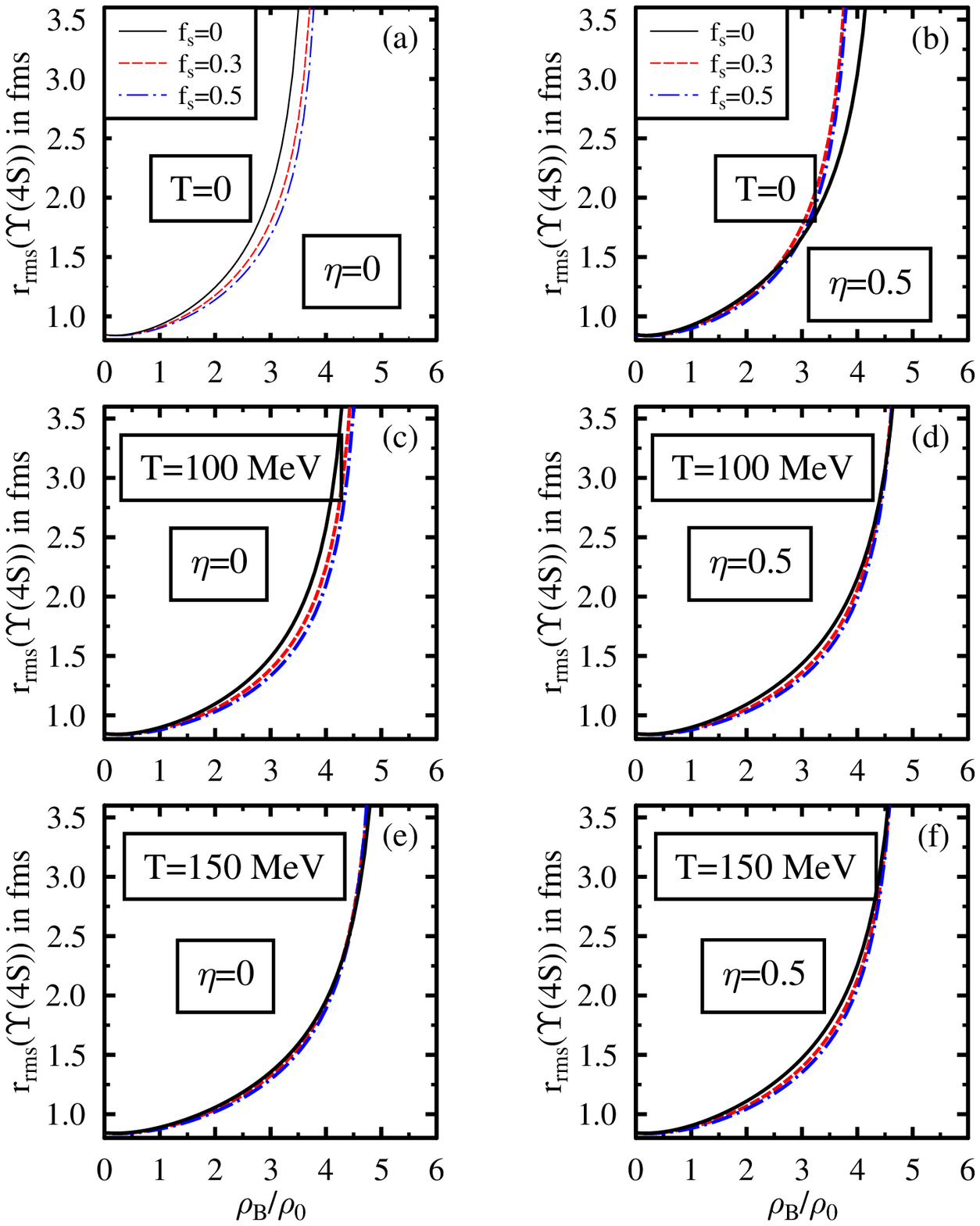}
\caption{(Color online)
The rms radius of $\Upsilon(4S)$ plotted as a function of the
baryon density in units of nuclear matter saturation density
at given temperatures,
for different values of the 
strangeness fraction for the isospin symmetric ($\eta$=0) as well as isospin 
asymmetric ($\eta$=0.5) hadronic matter.
}
\label{rmsupsln4s}
\end{figure}

In this section, we investigate the effects of density and temperature 
on the masses of the bottomonium states for given isospin-asymmetry 
and strangeness of the hadronic medium. Within a chiral SU(3) model,
the scalar gluon condensate in the hadronic matter has been calculated 
from the change in the dilaton field and scalar fields. Using the leading
order QCD formula for the mass shift of heavy quarkonium state,
the medium modification of the charmonium states $J/\psi$,
$\psi(3686)$ and $\psi (3770)$ were studied from the 
modification of the gluon condensate in the hadronic matter \cite{leeko,amepja}.
In the present work, we investigate the medium modification of the
$\Upsilon$-states using equation (\ref{mass1}) from the
value of the scalar gluon condensate calculated in the chiral SU(3) model. 
The value of the mass of the b-quark
is taken to be $m_b$=5.36 GeV \cite{repko} in the present investigation. 
This value of $m_b$ is similar
to the value of 5.1 GeV \cite{bhanotpeskin9} and 5.005 GeV 
\cite{spmmesonspect} used in the potential model calculations for the study
of the bottomonium spectroscopy. As has already been
mentioned the wave functions of the $\Upsilon$-states have been
taken to be harmonic oscillator eigenfunctions as given by equation
 (\ref{wavefn}) 
and the values of the harmonic 
oscillator strength, $\beta$ for the states are determined from the
experimental values of the leptonic decay widths of the bottomonium states,
given by equation (\ref{leptonicdw}). The  values of the decay widths 
($\Upsilon \rightarrow e^+e^-$) 
of 1.34 keV, 0.612 keV, 0.443 keV and 0.272 keV
for the $\Upsilon$-states, $\Upsilon (1S)$, $\Upsilon (2S)$,
$\Upsilon (3S)$ and $\Upsilon (4S)$ \cite{pdg2012}
give the values of the harmonic 
oscillation strength, $\beta$ as 1309.2, 
915.4, 779.75 and 638.6 MeV respectively. These values of the parameter
$\beta$ may be compared with the calculations of the heavy quarkonium systems
using a quark model \cite{spmmesonspect} with a linear confining potential, 
calculated as 1291 and 953.46 MeV for the $\Upsilon (1S)$ and $\Upsilon (2S)$
states.
  
In figure \ref{ggscdens}, the density dependence of the
scalar gluon condensate, 
$G_0=\langle (\alpha_s/{\pi})G_{\mu \nu}^aG^{\mu \nu a}\rangle$ 
is shown for temperatures, T=0, 100 and 150 MeV,
for the isospin symmetric matter in subplots (a), (c) and (e),
for values of the strangeness fraction, $f_s$ as 0, 0.3 and 0.5.
These are compared with the values for the isospin 
asymmetric matter with $\eta$=0.5 plotted in (b), (d) and (f).
The scalar gluon condensate is calculated in the chiral SU(3) model,
using equations (\ref{chiglu}) and (\ref{lsb}), from the in-medium values 
of the dilaton field, $\chi$ and the scalar fields, 
$\sigma$, $\zeta$ and $\delta$ in the isospin asymmetric strange hadronic
medium, obtained by solving the coupled equations of motion of these fields.  
For the isospin symmetric nuclear matter, the value of the scalar gluon
condensate 
changes from the vacuum value of 1.9387$\times$10$^{-2}$GeV$^4$ to 
1.9$\times$10$^{-2}$GeV$^4$ at the nuclear matter saturation density,
and to 1.822$\times$10$^{-2}$GeV$^4$ and 
1.7155$\times$10$^{-2}$GeV$^4$
at densities 2$\rho_0$ and 4$\rho_0$ respectively.
In the presence of hyperons in the system, the dilaton field is 
observed to have a larger drop in the medium. When the quark masses are neglected,
the second term of the trace of the energy momentum tensor does not contribute
and the scalar gluon condensate becomes proportional to the fourth power
of the dilaton field. However, the contributions
from the quark condensate term in the trace of the energy momentum
tensor expressed in terms of the scalar fields, $\sigma$ and
$\zeta$, modifies the value of the scalar gluon condensate in the medium.
At zero temperature,  for symmetric hadronic matter,
the modification arising from the finite strangeness fraction 
is observed to be small, with the value of the scalar gluon condensate
amounting to 1.9075$\times$10$^{-2}$GeV$^4$,
1.8336$\times$10$^{-2}$GeV$^4$,
and 1.7223$\times$10$^{-2}$GeV$^4$,
for $f_s$=0.3 and 
1.9117$\times$10$^{-2}$GeV$^4$,
1.8412$\times$10$^{-2}$GeV$^4$,
and 1.7248$\times$10$^{-2}$GeV$^4$,
for $f_s$=0.5,
for densities of $\rho_0$, 2$\rho_0$ and 4$\rho_0$ respectively.
At T=0, the effect from the finite $f_s$ is observed to be 
larger for the isospin asymmetric case as compared to the symmetric 
situation. The difference in the in-medium gluon condensate from the 
vacuum value is observed to decrease as the temperature is increased,
due to the fact that the drop in the magnitude in the dilaton and the
scalar fields, $\sigma$ and $\zeta$ decrease with increase in temperature.

In figures \ref{dmupsln1s}, \ref{dmupsln2s}, \ref{dmupsln3s} and 
\ref{dmupsln4s},
we show the density dependence of the mass shifts of the upsilon states, 
$\Upsilon(1S)$,
$\Upsilon(2S)$, $\Upsilon(3S)$ and $\Upsilon(4S)$ respectively,
in isospin asymmetric strange hadronic matter. The mass of $\Upsilon (1S)$
is observed to drop from its vacuum value of 9460.3 MeV to 9459.94 MeV
(9458.12 MeV) at density of 
$\rho_0$ (4$\rho_0$) for symmetric nuclear matter at zero temperature. 
For the asymmetric nuclear matter ($\eta$=0.5),    
the in-medium mass is changed to 9459.956 MeV (9458.36 MeV) 
at density $\rho_0$ (4$\rho_0$). 
The strangeness fraction in the medium is seen to lead to marginal modification
for the symmetric matter, whereas it is larger for the asymmetric situation
for T=0.
However, the value of the mass shift remains small, of the order of about 
3 MeV at a density
of six times nuclear matter density. With increase in temperature, the
drop is seen to be even less (of the order of 2.5 MeV at a density of
6$\rho_0$). For $\Upsilon (2S)$
plotted in figure \ref{dmupsln2s}, the mass 
is observed to be modified to 10019.83 MeV (10002.49 MeV) 
at $\rho_0$ (4$\rho_0$) from its vacuum value
of 10023.26 MeV. However, the effects from temperature, strangeness and 
isospin asymmetry remain small as compared to the modification of the mass
from density effects.
For $\Upsilon (3S)$ shown in figure \ref{dmupsln3s}, the mass
shift is observed to be 12.21 MeV (73.916 MeV) from its vacuum value of
10355.2 MeV,
at density of $\rho_0$ (4$\rho_0$) for symmetric nuclear matter at T=0.
The dependence on $f_s$ is seen to be larger for the asymmetric case
at high densities. The mass shift of the $\Upsilon$ state is proportional
to the difference in the value of the scalar gluon condensate from the
vacuum value, with the constant of proportionality determined by 
the integral of equation (\ref{mass1}) whose integrand is given 
in terms of the wave function of the specific $\Upsilon$ state.
For given values of the temperature, isospin asymmetry, strangeness fraction
and density of the medium, the ratio of the magnitudes of the mass shifts for 
the $\Upsilon(1S)$, $\Upsilon (2S)$, $\Upsilon (3S)$ and $\Upsilon (4S)$
states turns out to be the ratio of the magnitudes of the integral 
calculated from their respective wave functions.
These mass shifts are observed to be in the ratio 
$\Delta m_{\Upsilon (1s)}$:$\Delta m_{\Upsilon (2s)}$:$\Delta m_{\Upsilon (3s)}$:$\Delta m_{\Upsilon (4s)}$ = 1$\;$:$\;$9.53$\;$:$\;$33.9$\;$:$\;$137.8.
The mass drop of the excited states are thus observed to be larger 
for the excited states and for 
$\Upsilon (4S)$, the drop in the mass is seen to be about 49.64 MeV
(300.4 MeV) from the vacuum mass of 10579.4 MeV, at a density of 
$\rho_0$(4$\rho_0$) for symmetric nuclear matter at zero temperature.
We might note here that the mass of the $\Upsilon$ state has initially
an increase with density, as the contribution from the finite quark mass 
term dominates over the first term in the expression for the scalar gluon
condensate given by equation (\ref{chiglu}). The rise is observed to be
about 3.4 MeV at a density of about 0.2$\rho_0$ for $\Upsilon (4S)$,
followed by a drop in the mass as the density is further increased.
In the case when the finite quark mass term in the trace of the 
energy momentum tensor in QCD is neglected, there is seen to be
a monotonic drop of the $\Upsilon$ states with increase in density,
since the dilaton field dereases with density.
 
With the harmonic oscillator parameter, $\beta$ of the wave function 
of the bottomonium states as fitted from their leptonic decay widths,
the root mean square radii for the states, $\Upsilon(1S)$,
$\Upsilon(2S)$, $\Upsilon (3S)$ and  $\Upsilon (4S)$ are obtained as
0.1843, 0.4027, 0.5928 and 0.8466 fermis respectively in the present 
investigation. 
These values are similar to the values  for the rms radii obtained from using a
Cornell potential for the bottomonium bound state of Ref. \cite{eichten80},
of 0.2, 0.48, 0.72 and 0.92 fermis for these  bottomonium states and,
0.1869 and 0.3865 fermis for the $\Upsilon (1S)$ and $\Upsilon (2S)$ states
calculated in a quark model \cite{spmmesonspect} using a confining
linear potential. In the hadronic medium, due to the
mass drop in the bottomonium states, the strength of the harmonic oscillator 
wave function, $\beta$ is modified from which we can obtain an estimate for the size
of the bottomonium state in the hadronic medium. For the states, $\Upsilon(1S)$,
$\Upsilon (2S)$, $\Upsilon (3S)$ and $\Upsilon(4S)$, the change in the 
strength of the quarkonium wave function is obtained \cite {leeko,amepja}
from $\Delta \beta^2=\frac{ 2}{3}M_\Upsilon\Delta M_{\Upsilon}$,
$\Delta \beta^2=\frac{ 2}{7}M_\Upsilon\Delta M_{\Upsilon}$,
$\Delta \beta^2=\frac{ 2}{11}M_\Upsilon\Delta M_{\Upsilon}$,
$\Delta \beta^2=\frac{ 2}{15}M_\Upsilon\Delta M_{\Upsilon}$,
for the states $\Upsilon(1S)$, $\Upsilon(2S)$, $\Upsilon(3S)$,
and $\Upsilon(4S)$ respectively. The rms radii of these states
in the isospin asymmetric hot strange hadronic matter are plotted
in figures \ref{rmsupsln1s}, \ref{rmsupsln2s}, \ref{rmsupsln3s}
and \ref{rmsupsln4s}. There is seen to be increase in the rms radii
of these states with density and this rise is seen to be
especially  prominent for $\Upsilon(4S)$ state. These can have
consequences on the scattering cross-sections of these states 
by the baryons in the medium. This is because the leading order 
QCD calculations \cite{pes1,oh} show that
this scattering cross-section is  proportional to the root mean square radii, 
$r_{rms}^2$ \cite{frimanlee}. 
Hence the increase in the size of the $\Upsilon$-states in the medium 
can enhance their decay widths.
 
\section{Summary}
In the present work, we have  investigated the mass modification of the
bottomonium states ($\Upsilon (1S)$, $\Upsilon (2S)$, $\Upsilon (3S)$ and
$\Upsilon (4S)$), using the leading order QCD mass formula, from the medium 
modification of the scalar gluon condensate. The gluon condensate 
in the isospin asymmetric strange hadronic matter is calculated in a chiral 
SU(3) model. The broken scale invariance of QCD is incorporated 
into the hadronic model by introducing a scalar dilaton field. 
In the limit of massless quarks, the scalar gluon condensate
is proportional to the fourth power of the dilaton field. However, in the
case of finite quark masses, there is contribution to the scalar gluon
condensate from the quark condensates, which are determined in the chiral
SU(3) model from the values of the scalar fields, $\sigma$, $\zeta$
and $\delta$ in the hadronic medium. The mass shifts of the $\Upsilon$-states
are observed to be larger for the excited states and are of the order of
few hundred MeV for $\Upsilon (4S)$ state. The density effect is the dominant
medium effect as compared to the effects from strangeness fraction, isospin 
asymmetry and the temperature of the hadronic matter. These mass shifts 
can possibly show in the dilepton spectra arising from the compressed 
baryonic matter in the future facility at GSI, when there is access
to higher energies as compared to the planned energy range 
at CBM. The density effects of the mass modifications
of the bottomonium states should also show up in the dilepton 
spectra at SPS. The $\Upsilon$ states have already been 
measured in pA collisions at incident energy of 450 AGeV
($\sqrt s$=29.1GeV) by the NA50 Collaboration \cite{pANA50bottomonium},
which can provide a baseline for the $\Upsilon$ 
production studies to be carried out in the ion-ion collisions
at higher centre of mass energies at SPS. In the present work,
the medium modifications of rms radii of the $\Upsilon$ states have been 
calculated due to the change in the strength in the harmonic oscillator 
wave functions of these states in the hadronic medium. There is seen 
to be an increase in the rms radii of these states with density 
in the present investigation. 
This can lead to appreciable contribution to the decay widths of the 
bottomonium states due to scattering from the nucleons in the hadronic
medium.

\acknowledgements
AM would like to thank S.P.Misra and H. Mishra for discussions and
Department of Science and Technology, Government of India  
(project number SR/S2/HEP-031/2010) for financial support. 
DP acknowledges financial support from University of Grants Commission,
India (Sr. No. 2121051124, Ref. No. 19-12/2010(i)EU-IV).


\end{document}